\documentclass[12pt]{iopart}

\newtheorem{theorem}{Theorem}
\newtheorem{lemma}{Lemma}

\usepackage{braket}
\usepackage{bm}

\usepackage{color}

\eqnobysec

\usepackage{graphicx}
\usepackage{latexsym}
\def\qed{\hfill $\Box$} 

\begin{document}

\title[Properties of Deformed Fredkin Spin Chain]{Finite-size Gap, Magnetization, and Entanglement of Deformed Fredkin Spin Chain}
\author{Takuma Udagawa and Hosho Katsura}

\address{Department of Physics, Graduate School of Science, The University of Tokyo, Hongo, Bunkyo-ku, Tokyo 113-0033, Japan}
\eads{\mailto{udagawa@cams.phys.s.u-tokyo.ac.jp} and \mailto{katsura@phys.s.u-tokyo.ac.jp}}
\vspace{10pt}
\begin{indented}
\item \today
\end{indented}

\begin{abstract}

We investigate ground- and excited-state properties of the deformed Fredkin spin chain proposed by Salberger, Zhang, Klich, Korepin, and the authors. This model is a one-parameter deformation of the Fredkin spin chain, whose Hamiltonian is $3$-local and translationally invariant in the bulk. The model is frustration-free and its unique ground state can be expressed as a weighted superposition of colored Dyck paths.  
We focus on the case where the deformation parameter $t>1$.
By using a variational method, we prove that the finite-size gap decays at least exponentially with increasing the system size. We prove that the magnetization in the ground state is along the $z$-direction, namely $\braket{s^x}=\braket{s^y}=0$, and show that the $z$-component $\braket{s^z}$ exhibits a domain-wall structure. 
We then study the entanglement properties of the chain. In particular, we derive upper and lower bounds for the von Neumann and R\'enyi entropies, and entanglement spectrum for any bipartition of the chain.
\end{abstract}


%
%
%
%
%

\section{Introduction}

Spin chains play an important role in many branches of physics~\cite{Bonner, Affleck_review, Korepin_book, Kitanine, Calabrese, Beisert}. In particular, solvable and/or integrable cases have received much attention~\cite{Bethe, LSM, McCoy, Jimbo-Miwa, AKLT1, AKLT2}. One of the topics discussed in the field of spin chains is  the von Neumann (entanglement) entropy of ground state. For gapless spin chains with local interactions, it is typical that the entanglement entropy of ground state scales logarithmically with the system size \cite{gaplessCFT}. For gapped spin chains with local interactions, Hastings proved that the entanglement entropy obeys an area law, i.e., the entanglement entropy is bounded from above by a constant independent of the system size \cite{hastings}. Of course, ground states with a higher degree of entanglement are realized if strongly inhomogeneous \cite{Latorre, rainbow} and/or non-local interactions are allowed. 
We stress, however, that locality of interactions is important in many models because most quantum many-body systems can be well approximated by local Hamiltonians. 

Recently, a new class of exactly solvable spin chains has been proposed. They are deeply related to combinatorial problems such as the enumeration of lattice paths. One example is an integer-spin chain studied in~\cite{shor, shor2}. The Hamiltonian of this model is 2-local and translationally invariant in the bulk. The exact ground state can be expressed as a uniform superposition of colored Motzkin paths by identifying paths with spin configurations. Many properties of the ground state, including entanglement and magnetization, have been studied~\cite{shor, shor2, mova}. Further, it was shown that the model exhibits critical behavior, i.e., the spectral gap scales polynomially with the inverse of the system size. A one-parameter deformation of the model was introduced and studied in~\cite{klich}. It was found that the model undergoes a quantum phase transition when the deformation parameter $t$ is varied. The exponential gap scaling of the model at $t>1$ was proved in \cite{dMotGap}. Furthermore, some topological properties of the model were studied numerically in \cite{Haldane}.

Another model which has a similar combinatorial interpretation is a Fredkin spin chain \cite{fred, violation}. The model describes half-integer spins and a local interaction among three neighboring spins. The interaction is frustration-free, i.e., the ground state minimizes each interaction term. It has been proved that the unique ground state of the Fredkin chain is a uniform superposition of colored Dyck paths. Note that in terms of the number of colors $s=1, 2, ...$, the spin quantum number is expressed as $s-\frac{1}{2}$. 
For the colorless (spin-$\frac{1}{2}$) case, the entanglement entropy of the ground state scales as $\log{n}$ where $n$ is half the system size.\footnote{
Although the entanglement scaling obeys a logarithmic law as in conformal field theories \cite{Calabrese, Wilczek1, Wilczek2}, the model in the continuous limit is not described by any conformal field theory. In fact, the dynamical critical exponent in the model was found to be $z=3.2$, which is far from the conformal value $z=1$ \cite{Fradkin}.
}
When the color degrees of freedom are included,  it scales as $\sqrt{n}$. 
In either case, the ground state does not obey an area law. This together with the contraposition of Hastings' theorem~\cite{hastings} implies that the model is gapless for all $s$. In fact, the finite-size gap of the Fredkin chain is polynomially small in the system size, as proved by Movassagh in Ref. \cite{fredgap}. 
Later, in Ref. \cite{qfred}, a one-parameter deformation of the Fredkin chain whose ground state is a weighted superposition of colored Dyck paths was proposed. The deformation is characterized by $t$ and the original Fredkin chain corresponds to the case $t=1$.
The behavior of the entanglement entropy of half chain depends on both $t$ and the number of colors $s$ in a complicated way. When $t<1$, the entanglement entropy obeys an area law, irrespective of the value of $s$.  
Similarly, the entanglement entropy for $t>1$ and $s=1$ also obeys an area law. 
On the other hand, when $t>1$ and $s>1$, the entanglement entropy is tightly bounded by $n\log{s}$, implying the volume-law scaling and high entanglement. Therefore, the model with $s>1$ exhibits a quantum phase transition from an area-law to a volume-law entanglement in the ground state.

In this paper, we study various properties of the deformed Fredkin chain, particularly focusing on the case $t>1$. The first is a spectral gap which is identical to the energy of the first excited state since 
the ground-state energy is zero. The contraposition of Hastings' theorem~\cite{hastings} again proves that the model is gapless when $t>1$ and $s>1$.  
However, this argument alone does not specify how fast the finite-size gap decays with the system size. Furthermore, this does not tell us whether the system is gapped or gapless for the colorless case ($s=1$) with $t>1$. In this work, by a variational method,  
we obtain an upper bound for the spectral gap and prove that the finite-size gap $\Delta E$ is at most exponentially small in the system size $2n$, namely, $\Delta E\leq \Or(t^{-2n})$.

The second is the magnetization. We show that the magnetization is along the $z$-direction at any site. In addition, we find that the $z$-component of the magnetization, say $\braket{s^z}$, exhibits a domain-wall structure, in which $\braket{s^z} \sim s/2$ ($-s/2$) holds for almost all sites in the left- (right-) half system.  

Thirdly, we study various entanglement properties. The entanglement entropy of half of the chain in the deformed Fredkin chain was evaluated in \cite{qfred}. In this work, we present a more thorough analysis on the entanglement. We study the Schmidt rank, entanglement and R\'enyi entropies for any bipartition of the chain into two subsystems. 
We derive upper and lower bounds for the entanglement entropy and show that the ground state for $t>1$ obeys an area law when $s=1$, while it obeys a volume law when $s>1$. 
We also calculate the entanglement spectrum and show for large system sizes that the entanglement energy per site is equidistant at low energies, which is reminiscent of the spectrum of the harmonic oscillator.

The rest of the paper is organized as follows:  In section 2, we explain the model and its ground state. We then give a lemma and a theorem for the Schmidt coefficients which will be used frequently in the sequel. 
In section 3, we prove that the spectral gap is upper bounded by $\Delta E\leq \Or(t^{-2n})$. 
In section 4, we show that the magnetization is along the $z$-direction, and estimate the expectation value of $\hat{s}^z$. In section 5, we evaluate the Schmidt rank, entanglement and R\'enyi entropies, and entanglement spectrum for any bipartition of the chain. Section 6 is the conclusion. 
In appendix, the estimates of some constants used in the main text are shown.

\section{Hamiltonian and ground state}
Here we explain the deformed Fredkin spin chain which was introduced in Ref.~\cite{qfred}.

\subsection{Colored Dyck paths}
\begin{figure}
\begin{center}
\includegraphics[width=7cm]{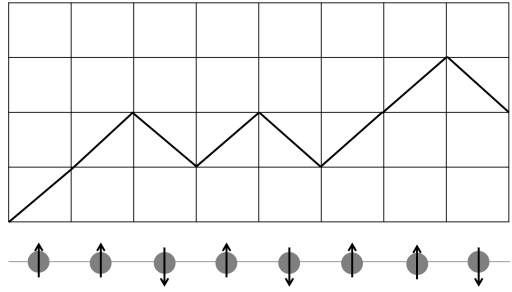}
\caption{
An example of the relation between Dyck paths and spin configurations. The bottom represents the state
$|\uparrow\uparrow\downarrow\uparrow\downarrow
\uparrow\uparrow\downarrow\rangle$.
}
 \label{notdyck}
\end{center}
\end{figure}

The deformed Fredkin spin chain is deeply related to lattice paths. 
For clarity, we first consider the spin-$\frac{1}{2}$ case corresponding to the colorless case. We regard local up- and down-spin states as $(1,1)$ and $(1,-1)$ steps, respectively. In the following, we call 
$(1,1)$ and $(1,-1)$ as up ($\nearrow$) and down ($\searrow$) steps, respectively.\footnote{The notations $\{ \uparrow, \downarrow \}$ and $\{\nearrow, \searrow \}$ are used interchangeably.}   
With this identification, any spin configuration maps to the corresponding lattice path (as an example see figure \ref{notdyck}). To treat more general cases, we introduce color degrees of freedom. For the spin-($s-\frac{1}{2}$) case, we assign a color $c_j$ ($c_j=1,2,...,s$) to the state whose $s^z$ component is $\pm$($c_j-\frac{1}{2}$). Then we identify a local spin $\pm(c_j-\frac{1}{2})$ state with a $(1,\pm1)$ step with color $c_j$. With this identification, any spin configuration in the spin-($s-\frac{1}{2}$) chain maps to the corresponding colored path. 

\begin{figure}
\begin{center}
\includegraphics[width=7cm]{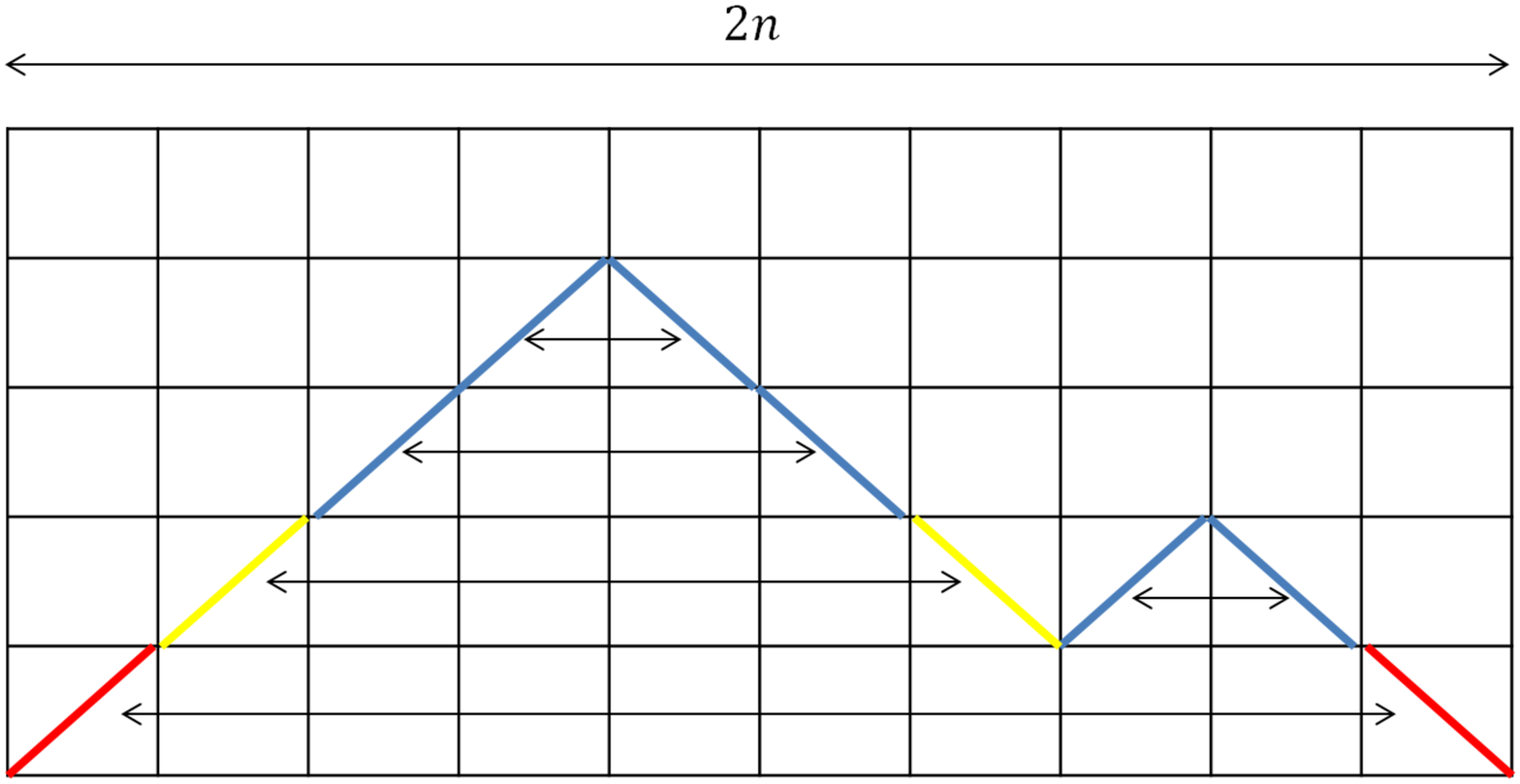}
\caption{An colored Dyck path with $s=3$ colors on a chain of length $2n=10$.} \label{dyck}
\end{center}
\end{figure}

We now introduce the notion of colored Dyck paths. They are a subset of colored paths that connect the coordinates $(0,0)$ and $(2n,0)$. More precisely, the definition is as follows. 
A path on $2n$ steps is called a colored Dyck path iff (i) it starts at $(0,0)$ and ends at $(2n,0)$ with steps $(1,1)$ and $(1,-1)$, (ii) it never goes below the $x$-axis, and (iii) matched steps in the path have the same color. An example of colored Dyck paths is shown in figure \ref{dyck}. Note that we say that a pair of up and down steps is {\it matched} when these steps face each other, such as those connected by arrows shown in figure \ref{dyck}. The bijection between colored paths and spin configurations maps each colored Dyck path to the corresponding spin configuration.

\subsection{Hamiltonian and unique ground state}
The Hamiltonian of the deformed Fredkin spin chain of length $2n$ \cite{qfred} is
\begin{eqnarray}
H(s,t)=H_{F}(s,t)+H_{X}(s)+H_{\partial}(s),
\end{eqnarray}
where
\begin{eqnarray}
H_F(s,t)=\sum_{j=2}^{2n-1}\sum_{c_1,c_2,c_3=1}^{s}\Bigl(\ket{\phi_{j,+}^{c_1,c_2,c_3}}\bra{\phi_{j,+}^{c_1,c_2,c_3}}+\ket{\phi_{j,-}^{c_1,c_2,c_3}}\bra{\phi_{j,-}^{c_1,c_2,c_3}}\Bigr)
\end{eqnarray}
with
\begin{eqnarray}
\ket{\phi_{j,+}^{c_1,c_2,c_3}}=\frac{1}{\sqrt{1+t^2}}\Bigl(\ket{\uparrow_{j-1}^{c_1}\uparrow_{j}^{c_2}\downarrow_{j+1}^{c_3}}-t\ket{\uparrow_{j-1}^{c_2}\downarrow_{j}^{c_3}\uparrow_{j+1}^{c_1}}\Bigr),\\
\ket{\phi_{j,-}^{c_1,c_2,c_3}}=\frac{1}{\sqrt{1+t^2}}\Bigl(\ket{\uparrow_{j-1}^{c_1}\downarrow_{j}^{c_2}\downarrow_{j+1}^{c_3}}-t\ket{\downarrow_{j-1}^{c_3}\uparrow_{j}^{c_1}\downarrow_{j+1}^{c_2}}\Bigr),
\end{eqnarray}
and
\begin{eqnarray}
H_X(s)&=&\sum_{j=1}^{2n-1}\Biggl[\sum_{c_1\neq c_2}\ket{\uparrow_{j}^{c_1}\downarrow_{j+1}^{c_2}}\bra{\uparrow_{j}^{c_1}\downarrow_{j+1}^{c_2}}\\
&\,&+\frac{1}{2}\sum_{c_1,c_2=1}^{s}\Bigl(\ket{\uparrow_{j}^{c_1}\downarrow_{j+1}^{c_1}}-\ket{\uparrow_{j}^{c_2}\downarrow_{j+1}^{c_2}}\Bigr)\Bigl(\bra{\uparrow_{j}^{c_1}\downarrow_{j+1}^{c_1}}-\bra{\uparrow_{j}^{c_2}\downarrow_{j+1}^{c_2}}\Bigr)\Biggr],
\end{eqnarray}
\begin{eqnarray}
H_{\partial}(s)=\sum_{c=1}^{s}\Bigl(\ket{\downarrow_{1}^{c}}\bra{\downarrow_{1}^{c}}+\ket{\uparrow_{2n}^{c}}\bra{\uparrow_{2n}^{c}}\Bigr).
\end{eqnarray}
Here, the spin $\pm (c-1/2)$ states at site $j$ are denoted by $\ket{\uparrow_j^{c}}$ and $\ket{\downarrow_j^{c}}$, respectively. 
Since the Hamiltonian is a sum of projectors, it is positive semi-definite.

This model is frustration-free and the unique ground state is given by
\begin{eqnarray}
\ket{{\rm GS}}=\frac{1}{\mathcal{N}}\sum_{w\in\{ s-{\rm colored\,Dyck\,paths}\}}t^{\frac{1}{2}\mathcal{A}(w)}\ket{w},
\label{gs}
\end{eqnarray}
where $\mathcal{N}$ is the normalization factor and $\{ s$-colored Dyck paths$\}$ refers to the set of colored Dyck paths (or equivalently, spin configurations) defined in the previous subsection. 
Here, $\mathcal{A}$ is the area between the path $w$ and the $x$-axis. 
Now let us see that the state (\ref{gs}) is a zero-energy state. Since the coefficient of $\ket{w_{1,\cdots ,j-2}}\ket{\uparrow_{j-1}^{c_1}\uparrow_{j}^{c_2}\downarrow_{j+1}^{c_3}}\ket{w'_{j+2,\cdots ,2n}}$ is $t$ times the coefficient of $\ket{w_{1,\cdots, j-2}}\ket{\uparrow_{j-1}^{c_2}\downarrow_{j}^{c_3}\uparrow_{j+1}^{c_1}}\ket{w'_{j+2,\cdots ,2n}}$ in Eq. (\ref{gs}), the sum of these states is annihilated by the projection $\ket{\phi_{j,+}^{c_1,c_2,c_3}} \bra{\phi_{j,+}^{c_1,c_2,c_3}}$. This occurs for any three consecutive sites and the same goes for $\ket{\phi_{j,-}^{c_1,c_2,c_3}}\bra{\phi_{j,-}^{c_1,c_2,c_3}}$. Therefore, $H_F(s,t)$ annihilates the ground state Eq. (\ref{gs}). 
The term $H_X$ ensures that matched steps have the same color and identical Dyck paths with different coloring have the same weight. Further, $H_{\partial}$ penalizes paths which go below the $x$-axis. Therefore, the state (\ref{gs}) is indeed a zero-energy ground state. It is also easy to check that the ground state is unique: any other superposition of paths violates at least one of the projectors and has non-zero energy.

\subsection{Schmidt decomposition of the ground state}
 
Let us first introduce some notation. Let $\mathcal{D}_{a,b} \subseteq \{ \nearrow, \searrow \}^{2a+b}$ be the set of paths which consist of $a$ pairs of up-down steps and $b$ extra up steps. More precisely, $w\in \mathcal{D}_{a,b}$ iff any initial segment of $w$ contains at least as many up steps as down steps, and the total number of up steps is $a+b$. Note that $\mathcal{D}_{n,0}$ is identical to the set of (uncolored) Dyck paths consisting of $2n$ steps. Next we define 
\begin{equation}
M_{2a+b,b}(t) := \sum_{w\in \mathcal{D}_{a,b}} t^{\mathcal{A}(w)}.
\end{equation}
The number $M_{2a+b,b}(t)$ is deeply related to Carlitz-Riordan's $q$-ballot number \cite{qballot1}, introduced a long time ago in Ref.~\cite{qballot2}.\footnote{The number $M_{2n,0}(t)$ can be expressed in terms of Carlitz-Riordan's $q$-Catalan number.} 
We have the following lemma.
\begin{lemma}
Let $P(k)$ be the partition function of $k$. When $t>1$,  $M_{2a+b,b}(t)$ satisfies the following inequality:
\begin{eqnarray}
t^{\frac{1}{2}(2a+b)^2-a^2}\leq M_{2a+b,b}(t)<C(t)t^{\frac{1}{2}(2a+b)^2-a^2},
\end{eqnarray}
where $C(t)$ is an $n$ independent constant given by
\begin{equation}
C(t) = \sum^\infty_{k=0} t^{-2k} P(k).
\end{equation}
\end{lemma}

\smallskip

\begin{figure}
\begin{center}
\includegraphics[width=7cm]{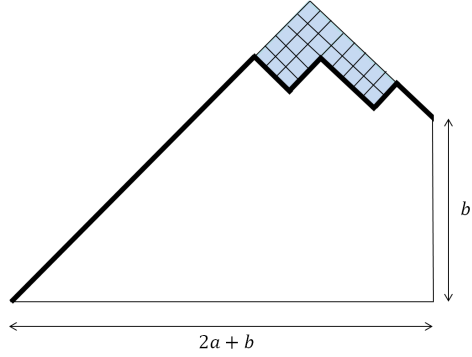}
\caption{
A path whose area is $\frac{1}{2}(2a+b)^2-a^2-2k$ corresponds to a Young diagram made up of $k$ boxes. The blue area shows an example. Since the region where we can make Young diagrams is restricted, the coefficient of $t^{\frac{1}{2}(2a+b)^2-a^2-2k}$ in Eq. (\ref{daaa}) is equal to or less than $P(k)$. } 
\label{nf}
\end{center}
\end{figure}

\noindent
$Proof$.\\
The proof of this lemma is very similar to that of Lemma 1 in \cite{qfred}. One can easily see that $M_{2a+b,b}(t)$ satisfies
\begin{eqnarray}
M_{2a+b,b}(t)&=&t^{\frac{1}{2}(2a+b)^2-a^2}+t^{\frac{1}{2}(2a+b)^2-a^2-2}+2t^{\frac{1}{2}(2a+b)^2-a^2-4}+\cdots +t^{a+\frac{1}{2}b^2} \label{daaa}\nonumber\\
&<&t^{\frac{1}{2}(2a+b)^2-a^2}\sum_{k=0}^{\infty}t^{-2k}P(k).
\end{eqnarray}
The coefficient of a term such as $t^{\frac{1}{2}(2a+b)^2-a^2-2k}$ in Eq. (\ref{daaa}) is the number of paths ($\in \mathcal{D}_{a,b}$) of area $\frac{1}{2}(2a+b)^2-a^2-2k$, and it is equal to or less than the partition function $P(k)$ (see figure \ref{nf}). 
Since the function $C(t)$ converges when $t>1$ \cite{pf}, we have
\begin{eqnarray}
t^{\frac{1}{2}(2a+b)^2-a^2}\leq M_{2a+b,b}(t)<C(t)t^{\frac{1}{2}(2a+b)^2-a^2}.
\end{eqnarray}
\qed

We turn to consider the Schmidt decomposition of the ground state. We divide the system into two parts by a cut between the sites $n_1$ and $n_1 +1$. Without loss of generality, we can assume $1\leq n_1 \leq n$ because the ground state is invariant under the combination of parity and spin-flip operations. 
The Schmidt decomposition of the ground state Eq. (\ref{gs}) takes the form
\begin{eqnarray}
\ket{{\rm GS}}=\sum_{m=0}^{n_1}\sqrt{p_{n_1,m}(s,t)}\sum_{x\in\{\uparrow^1,\uparrow^2,\cdots,\uparrow^s\}^m}\ket{\hat{C}_{0,m,x}}_{1,\cdots,n_1}\ket{\hat{C}_{m,0,\bar{x}}}_{n_{1}+1,\cdots,2n}, \label{gs2}
\end{eqnarray}
where
\begin{eqnarray}
p_{n_1,m}(s,t) =s^{-m}\, \frac{M_{n_1,m}(t) M_{2n-n_1,m}(t)}{M_{2n,0}(t)}, 
\label{ppp}
\end{eqnarray}
and $\ket{\hat{C}_{a,b,x}}$ is the area weighted superposition of spin configurations with $a$ excess $\downarrow$, $b$ excess $\uparrow$ and a particular coloring $x$ of unmatched arrows. 
$\bar{x}$ is the coloring which matches $x$. 
Note that $\ket{\hat{C}_{a,b,x}}$ are orthonormal.
From the above, it is obvious that $p_{n_1,m}(s,t)=s^{-m}p_{n_1,m}(1,t)$. A useful inequality for the Schmidt coefficients was proved for $n_1=n$ and $t>1$ in Ref.~\cite{qfred}. We generalize it to an arbitrary cut. 
In the following discussion, we write $N=n_1 -m$.
\begin{theorem}
When $t>1$, the Schmidt coefficients satisfy the following relation:
\begin{eqnarray}
p_{n_1,n_1-N}(s,t)=0  \qquad {\rm for~odd}~N
\label{odd}
\end{eqnarray}
and
\begin{eqnarray}
p_{n_1,n_1-N}(s,t)=\alpha(t,n_1,N)s^{N-n_1}t^{-\frac{1}{2}N^2-(n-n_1)N} \qquad {\rm for~even}~N,
\label{evennn}
\end{eqnarray}
with
\begin{eqnarray}
\frac{1}{C(t)}<\alpha(t,n_1,N)<C(t)^2. \label{dokaaan}
\end{eqnarray}
\end{theorem}

\smallskip

\noindent
$Proof$.\\
The equation (\ref{odd}) is obvious because there is no path that stops at $(n_1,\, n_1-N)$ for odd $N$. By applying Lemma 1 to Eq. (\ref{ppp}), one can easily see that $p_{n_1, n_1-N}(s,t)$ with $n$ even takes the form of Eq. (\ref{evennn}) and the coefficient $\alpha (t,n_1,N)$ satisfies the inequality (\ref{dokaaan}).
\qed

As the Schmidt coefficients are nonzero only for even $N$, we introduce $N'$ through $2N' = N$. 
Note that $\alpha(t,n_1 ,2N')$ satisfies $\sum_{N'=0}^{\lfloor \frac{n_1}{2} \rfloor}\alpha(t,n_1 ,2N') t^{-2N'^2-2(n-n_1)N' }=1$ because of the normalization condition of $p_{n_1,n_1-N}(1,t)$.
We will use Lemma 1 in section 3, and Theorem 1 in sections 4 and 5. 

\section{Finite-size gap}
In this section, we estimate the spectral gap $\Delta E$ by a variational method. Our trial state is
\begin{eqnarray}
\ket{\Phi}=\ket{\hat{C}_{0,2,x_1}}_{1,2,\cdots,2n},
\label{eq:trial}
\end{eqnarray}
where $\ket{\hat{C}_{0,2,x_1}}_{1,2,\cdots,2n}$ is the area-weighted superposition of spin configurations with two excess $\uparrow$ and the coloring $x_1 =( \uparrow^{1},\uparrow^{1} )\in \{\uparrow^1,\uparrow^2,\cdots,\uparrow^s\}^2 $. As any path contained in $\ket{\hat{C}_{0,2,x_1}}_{1,2,\cdots,2n}$ is not a colored Dyck path, $\ket{\Phi}$ is orthogonal to the ground state.

\begin{figure}
\begin{center}
\begin{tabular}{c}
\begin{minipage}{0.5\hsize}
\begin{center}
\includegraphics[width=11cm]{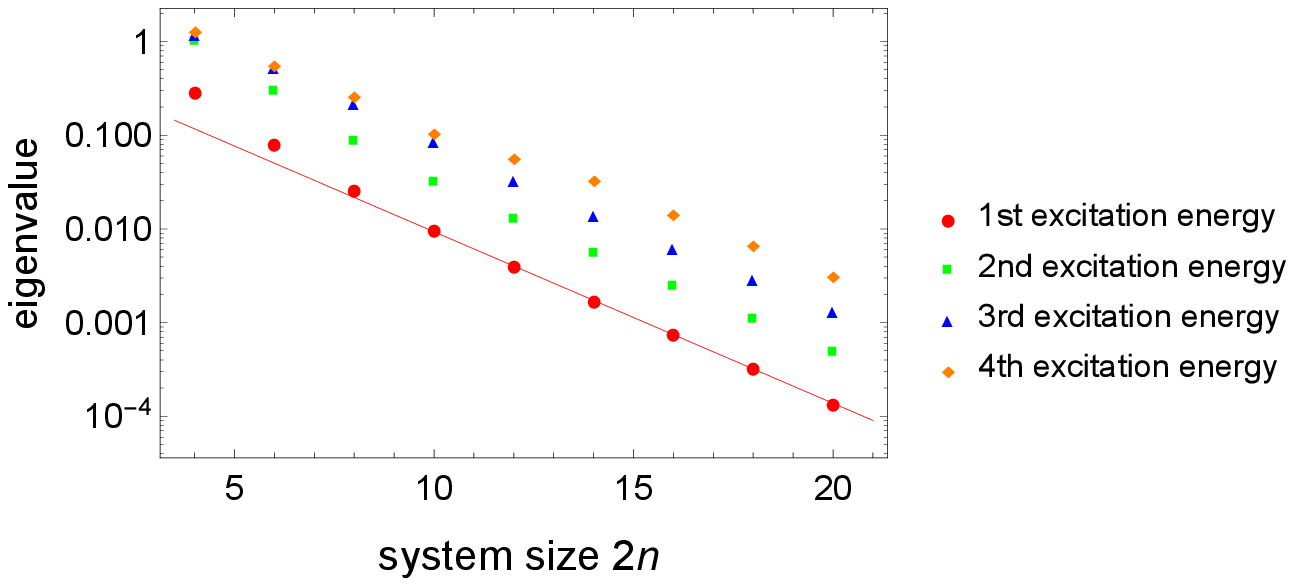}
\hspace{1.6cm}(a)\,\,$s=1$, $t=1.1$
\end{center}
\end{minipage}\\ \\
\begin{minipage}{0.5\hsize}
\begin{center}
\includegraphics[width=11cm]{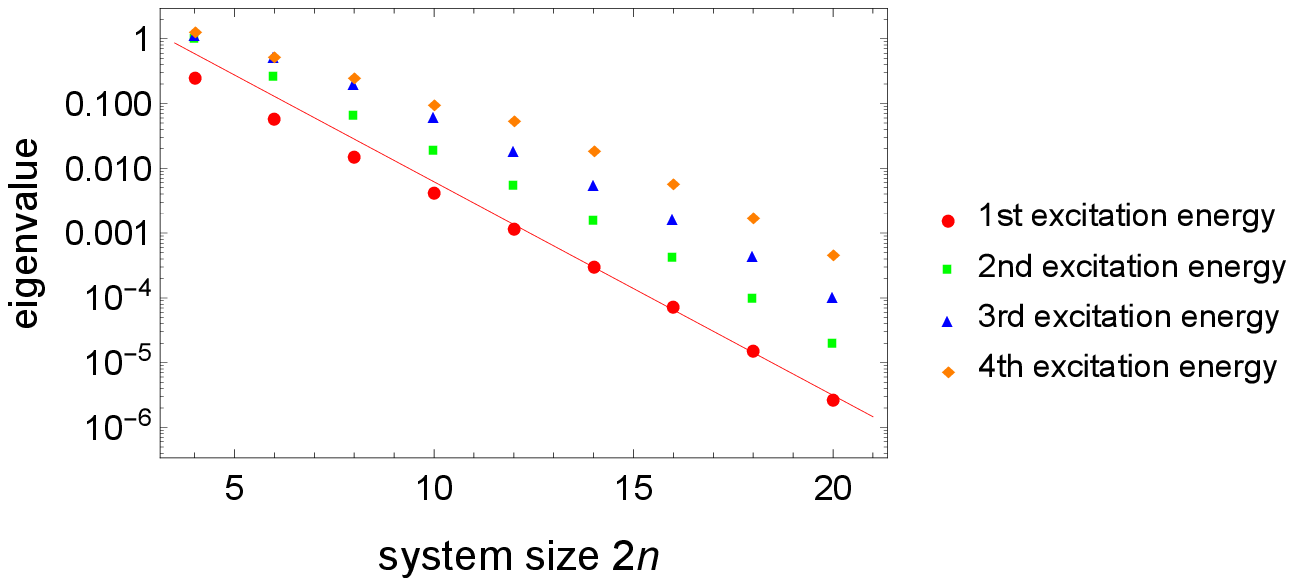}
\hspace{1.6cm}(b)\,\,$s=1$, $t=1.2$
\end{center}
\end{minipage}
\end{tabular}
\caption{The first few excitation energies as a function of system size for (a) $s=1$, $t = 1.1$ and (b) $s=1$, $t=1.2$. Note that the first excitation energy equals to the finite-size gap $\Delta E$ because the ground-state energy is zero. The red lines are fits to the data of $\Delta E$ for $2n = 12, 14, \cdots, 20$. From the fits, we find that $\Delta E = t^{-2n {\mathsf a}+{\mathsf b}}$ with ${\mathsf a}=4.42032$ and ${\mathsf b}=-4.85716$ for (a), and that with ${\mathsf a}=4.16493$ and ${\mathsf b}=13.7708$ for (b). This suggests that the closing of the gap is faster than ${\rm O}(t^{-2n})$, which is consistent with the bound Eq. (\ref{gapless}).}
\label{gapgap}
\end{center}
\end{figure}

The trial state Eq. (\ref{eq:trial}) is clearly a zero-energy state of the bulk Hamiltonian, i.e.,
\begin{eqnarray}
\Bigl[H_{F}(s,t)+H_{X}(s)\Bigr] \ket{\Phi}=0.
\end{eqnarray}
Therefore, the energy expectation value reads
\begin{eqnarray}
\bra{\Phi}H(s,t)\ket{\Phi}=\bra{\Phi}H_{\partial}(s)\ket{\Phi}=
\|\braket{\uparrow_{2n}^{1}|\Phi }\|^2 . 
\label{eve}
\end{eqnarray}
Note that only the paths which have $\uparrow^{1}$ at site $2n$ contribute to the energy. By considering the normalization factor of $\ket{\Phi}$ and using Lemma 1, one can see that equation (\ref{eve}) is upper bounded as follows:
\begin{equation}
\| \braket{\uparrow_{2n}^{1}|\Phi } \|^2=\frac{s^{n-1}M_{2n-1,1}(t)\, t^{\frac{3}{2}}}{s^{n-1}M_{2n,2}(t)}<C(t)\, t^{-2n+2}.
\label{gapless}
\end{equation}
This means that the finite-size gap is at most exponentially small in the system size $2n$ when $t>1$, namely $\Delta E\leq \Or(t^{-2n})$. However, this does not necessarily mean that the system is gapless in the infinite-size limit. In fact, we have two possibilities: (i) the ground state is degenerate and the system has a spectral gap, or (ii) the system is gapless. Our numerical results shown in figure {\ref{gapgap}} suggest the latter. Therefore, we expect that the deformed Fredkin chain at $t>1$ is gapless in the infinite-size limit. The numerical results for $s=1$ and $t>1$  also indicate that the power of the exponential gap scales linear with the system size, which is in accord with the behavior of the upper bound in Eq. (\ref{gapless}).

\smallskip

\noindent
$Remark$. The exponentially small spectral gap in the area-weighted Motzkin spin chain at $t>1$ was proved by Movassagh in Ref.~\cite{dMotGap}. The finite-size scaling of the gap is, however, different from ours. In \cite{dMotGap}, the finite-size gap is bounded from above by $\Delta E \leq 8nst^{-n^2/3}$, which decays much faster than ours. 

\section{Magnetization}
In this section, We calculate the magnetization in the ground state. Following the approach used in Ref.~\cite{mova}, we show that the expectations values of ${\hat s}^x_j$ and ${\hat s}^y_j$ at any site $j$ vanish. Then, we give a rough estimate for the expectation value of ${\hat s}^z_j$. 

\subsection{$\braket{s^x}$ and $\braket{s^y}$}
Let us first consider the  expectation value of $\hat{s}^x_j$. Using Eq. (\ref{gs}), the expectation value in the ground state can be expressed as
\begin{eqnarray}
\braket{s_{j}^{x}}&:=&\bra{{\rm GS}} \hat{s}_{j}^{x}  \ket{{\rm GS}}\nonumber\\
&=&\frac{1}{\mathcal{N}^2}\sum_{w,w'\in\{s-{\rm colored\,Dyck\,paths}\}}t^{\frac{1}{2}\mathcal{A}(w)+\frac{1}{2}\mathcal{A}(w')}\bra{w'} \hat{s}_{j}^{x} \ket{w}. 
\label{ssxx}
\end{eqnarray}
The key to the proof is that $\ket{w'}$ is orthogonal to ${\hat s}^x_j \ket{w}$. To see this, let us consider the action of ${\hat s}^x_j$ on local basis vectors. Let $\ket{{\mathsf s}, {\mathsf m}}_j$ (${\mathsf m}=-{\mathsf s}, ..., {\mathsf s}$) be a set of eigenstates of $({\hat {\bm s}}_j)^2$ and ${\hat s}^z_j$ at site $j$. We denote by ${\mathsf s}({\mathsf s}+1)$ and ${\mathsf m}$ their eigenvalues, i.e., $({\hat {\bm s}}_j)^2 \ket{{\mathsf s}, {\mathsf m}}_j={\mathsf s}( {\mathsf s}+1) \hbar^2 \ket{{\mathsf s}, {\mathsf m}}_j$ and $\hat{s}^z \ket{{\mathsf s}, {\mathsf m}}_j= {\mathsf m} \hbar \ket{{\mathsf s}, {\mathsf m}}_j$~\cite{sakurai}. Then, the operation of $\hat{s}^x_j$ on $\ket{{\mathsf s}, {\mathsf m}}_j$ reads
\begin{eqnarray}
\hat{s}^x_j \ket{{\mathsf s}, {\mathsf m}}_j=&&\frac{1}{2}\sqrt{{\mathsf s}({\mathsf s}+1)-{\mathsf m}({\mathsf m}+1)}\hbar\ket{{\mathsf s}, {\mathsf m}+1}_j\nonumber\\
&\,&+\frac{1}{2}\sqrt{{\mathsf s}({\mathsf s}+1)-{\mathsf m}({\mathsf m}-1)}\hbar\ket{{\mathsf s}, {\mathsf m}-1}_j .
\label{sx}
\end{eqnarray}
Since ${\hat s}^x_j$ changes the total ${\hat s}^z$ by one, $\hat{s}_{j}^{x} \ket{w}$ in equation (\ref{ssxx}) is no longer an $s$-colored Dyck path. Therefore, it follows from $\bra{w'} \hat{s}_{j}^{x} \ket{w}=0$ that
\begin{eqnarray}
\braket{s_{j}^{x}}=0. 
\label{sxn1}
\end{eqnarray}
In the same way, we see that the expectation value of $\hat{s}^y$ vanishes,
\begin{eqnarray}
\braket{s_{j}^{y}}=0. \label{syn1}
\end{eqnarray}
We note that equations (\ref{sxn1}) and (\ref{syn1}) are valid for any $t>0$.

\subsection{$\braket{s^z}$}

Before the calculation of $\braket{s_{j}^z}$, we give the following lemma. 
\begin{lemma}
Let $\braket{s_{j}^z}(s)$ be the expectation value of $\hat{s}^z_{j}$ in the ground state of the $s$-colored deformed Fredkin chain.  
At any site, $\braket{s_{j}^z}(s)$ for general $s$ and that for $s=1$ are related to each other thorough
\begin{eqnarray}
\braket{s_{j}^{z}}(s)=s\braket{s_{j}^{z}}(s=1).
\end{eqnarray}
\end{lemma}

\smallskip

\noindent
$Proof$.\\
In the ground state, the contribution of each color is the same weight. Therefore, $\braket{s_{j}^{z}}(s)$ is proportional to $\braket{s_{j}^{z}}(s=1)$ and the coefficient is 
\begin{eqnarray}
\frac{\frac{1}{2}+\frac{3}{2}+\cdots+(s-\frac{1}{2})}{s} \bigg( \frac{1}{2} \bigg)^{-1} =s.
\end{eqnarray}
\qed 

\smallskip

Because of Lemma 2, it suffices to consider the case $s=1$. Below we study the case where $s=1$ and $t>1$.

To calculate the expectation value of $\hat{s}^z$, we define the height operator $\hat{m}$ by
\begin{eqnarray}
\hat{m}_{j} :=2\sum_{i=1}^{j}\hat{s}_{i}^{z}.
\end{eqnarray}
Here we assume $j\leq n$. By using Theorem 1, $\braket{m_{j}}$ can be written as
\begin{eqnarray}
\braket{m_{j}}=\sum_{m=0}^{j} m\, p_{j,m}=j - \sum_{N'=0}^{\lfloor \frac{j}{2} \rfloor} 2N' \alpha(t,j,2N') t^{-2N'^2-2(n-j) N'}. \label{m}
\end{eqnarray}
If we take $j$ as $j =\gamma n$, where $\gamma \, (<1)$ is fixed, and sufficiently large $n$, then $\sum_{N'=0}^{\lfloor \frac{j}{2} \rfloor} 2N' \alpha(t,j,2N') t^{-2N'^2-2(n-j) N'}$ is much smaller than $j$. Hence, $\braket{m_{j}}$ is estimated as $\braket{m_{j}} \approx j$. Therefore the expectation value of $\hat{s}^z$ is
\begin{eqnarray}
\braket{s_{j}^{z}}(s=1) &=&\frac{1}{2}(\braket{m_{j}} -\braket{m_{j -1}})\approx \frac{1}{2}.
\end{eqnarray}

By using Lemma 2, we can also get the estimate for the general $s$ as 
\begin{eqnarray}
\braket{s_{j}^z}(s) \approx \frac{s}{2}.
\end{eqnarray}
Note that we have assumed $j \leq n$. If $j \geq n+1$, we have $\braket{s_{j}^z}(s) \approx -\frac{s}{2}$ because the ground state is invariant under the combination of parity and spin-flip operations. Therefore, the magnetization in the ground state exhibits a domain-wall structure, in accord with the intuition that the ground state at $t>1$ is dominated by a superposition of highest-area paths, $\ket{\uparrow^{c_1}_1 \cdots \uparrow^{c_n}_n \, \downarrow^{c_{n}}_{n+1} \cdots \downarrow^{c_1}_{2n} } $.

\begin{figure}
\begin{center}
\begin{tabular}{c}
\begin{minipage}{0.5\hsize}
\begin{center}
\includegraphics[width=11cm]{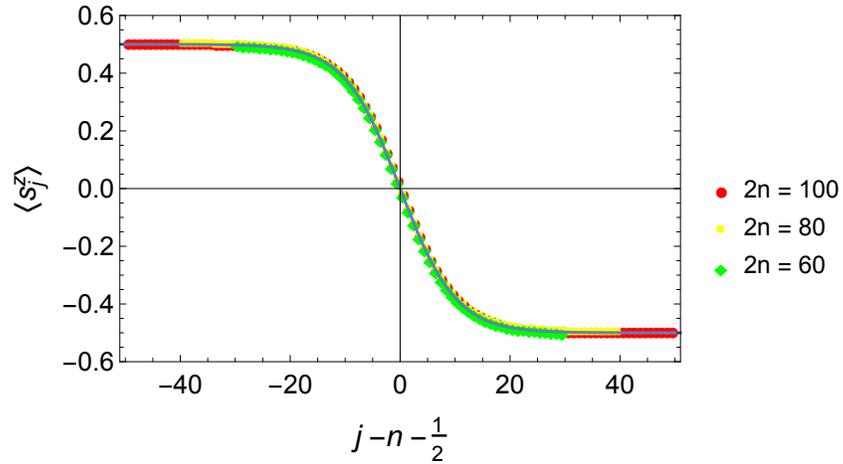}
\hspace{1.6cm}(a)\,\,$s=1$, $t=1.1$
\end{center}
\end{minipage}\\ \\
\begin{minipage}{0.5\hsize}
\begin{center}
\includegraphics[width=11cm]{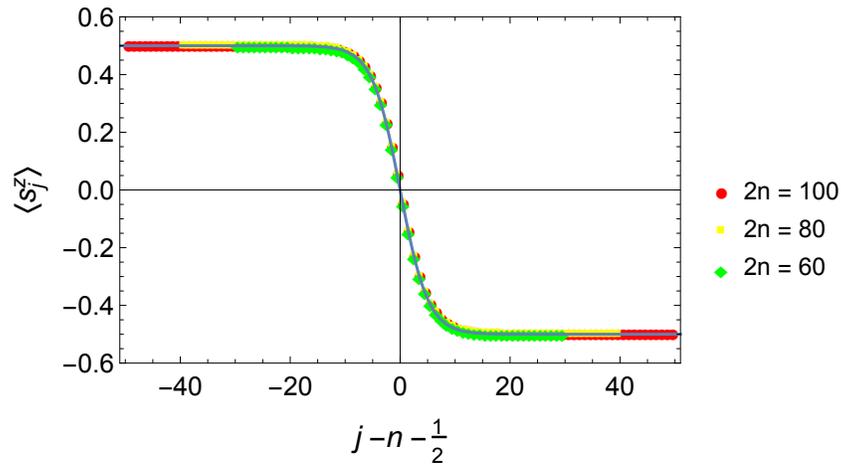}
\hspace{1.6cm}(b)\,\,$s=1$, $t=1.2$
\end{center}
\end{minipage}
\end{tabular}
\caption{Numerical results of $\braket{s_{j}^z}(s=1)$ for (a) $t=1.1$ and (b) $t=1.2$, on chains of length $2n=60$,$80$, and $100$. The data are fitted with a function of the form $\braket{s_{j}^z}(s=1)=-\frac{1}{2}\tanh [{\mathsf a}(j-n-\frac{1}{2})]$ (blue lines). 
From the fits, we find that 
${\mathsf a}=0.0990733$ for (a), and ${\mathsf a}=0.197042$ for (b).}
\label{szszsz}
\end{center}
\end{figure}

The numerical results of $\braket{s_{j}^z}(s=1)$ for $t=1.1$ and $1.2$ are shown in figure \ref{szszsz}. The profile of $\braket{s_{j}^z}$ around the center of the chain ($j=n$) is almost independent of the system size $2n$, as we can see from figure \ref{szszsz}. 
This suggests that the coefficient $\alpha$ in Eq. (\ref{m}) is almost independent of $n$ when $n-j$ is small. 
Assuming that this holds for any $t>1$, then the magnetization $\braket{s_{j}^z}$ as a function of the scaled position $\gamma=j/n$ becomes a step function in the infinite-size limit. 
This is consistent with our analysis above. 

\section{Entanglement properties}
In this section, we discuss the entanglement properties. To this aim, let us consider the reduced density matrix for the left block. 
From equation (\ref{gs2}) and Theorem 1,  we can derive it as
\begin{eqnarray}
\rho &=& {\rm Tr}_{(n_1 +1, \cdots , 2n)}\bigl[ \ket{{\rm GS}}\bra{{\rm GS}} \bigr] \nonumber\\
&=&\sum_{N'=0}^{\lfloor \frac{n_1}{2} \rfloor} \alpha s^{2N'-n_1} t^{-2N'^2 -2(n-n_1)N'} \nonumber\\
&\,&\,\,\,\,\,\,\,\,\,\,\,\,\,\,\,\, \times\sum_{x\in\{\uparrow^1,\uparrow^2,\cdots,\uparrow^s\}^{n_1 -2N'}}\ket{\hat{C}_{0,n_1 -2N',x}}\bra{\hat{C}_{0,n_1 -2N',x}}, 
\label{dondon}
\end{eqnarray}
where $\alpha$ is a short-hand notation for $\alpha(t,n_1 ,2N')$ in Eq. (\ref{evennn}).  

\subsection{Schmidt rank}
We can get the Schmidt rank $\chi_{n_1}(s)$ from equation (\ref{gs2}) and Theorem 1. When $s=1$, the Schmidt rank is
\begin{eqnarray}
\chi_{n_1}(s=1)=\left\lfloor \frac{n_1}{2} \right\rfloor +1. \label{schmidtrank1}
\end{eqnarray}
When $s>1$, the Schmidt rank is
\begin{eqnarray}
\chi_{n_1}(s>1)=\sum_{N'=0}^{\lfloor \frac{n_1}{2} \rfloor} s^{n_1 -2N'} = s^{n_1}\frac{1-s^{-2\bigl(\lfloor \frac{n_1}{2} \rfloor +1\bigr)}}{1-s^{-2}}.\label{schmidtrank2}
\end{eqnarray}

\subsection{Entanglement entropy}
The entanglement entropy is given by
\begin{eqnarray}
S_{n_1}(s,t)&=&-{\rm Tr}\bigl[ \rho \log \rho \bigr]\nonumber\\
&=&-\sum_{N'=0}^{\lfloor \frac{n_1}{2} \rfloor} \alpha t^{-2N'^2 -2(n-n_1)N' } \log\bigl[ \alpha s^{2N'-n_1} t^{-2N'^2 -2(n-n_1)N' } \bigr]\nonumber\\
&=&n_1 \log s \\
&\,&-2\log s \sum_{N'=0}^{\lfloor \frac{n_1}{2} \rfloor} \alpha N' t^{-2N'^2 -2(n-n_1)N' }  \label{ee1}\\
&\,&-\sum_{N'=0}^{\lfloor \frac{n_1}{2} \rfloor} \alpha t^{-2N'^2 -2(n-n_1)N' } \log\bigl[ \alpha t^{-2N'^2 -2(n-n_1)N' } \bigr].\label{ee2}
\end{eqnarray}
Here we used equation (\ref{dondon}) and the normalization condition $\sum_{N'=0}^{\lfloor \frac{n_1}{2} \rfloor} \alpha t^{-2N'^2 -2(n-n_1)N' } =1$. Now we turn to estimate the summations (\ref{ee1}) and (\ref{ee2}). First, the summation (\ref{ee1}) is bounded from above and below as follows:
\begin{eqnarray}
0&\geq& -2 \log s\sum_{N'=0}^{\lfloor \frac{n_1}{2} \rfloor} \alpha N' t^{-2N'^2 -2(n-n_1)N'}  \nonumber\\
&\geq& -2C(t)^2\log s \sum_{N'=0}^{\infty} N' t^{-2N'^2}  \, =: \,  D_1 (s,t).\label{ee3}
\end{eqnarray}
Here we have used Theorem 1 and the fact that $n_1 \leq n$. Next, we examine Eq. (\ref{ee2}). The key to obtaining an upper bound is the Gibbs inequality 
\begin{equation}
-\sum_{i}p_i \log p_i \leq -\sum_{i}p_i \log q_i,
\end{equation} 
which holds for any probability distributions $\{ p_i \}$ and $\{ q_i \}$ with equality iff $p_i = q_i$ for all $i$ \cite{gibbs}. By identifying $\{ p_i \}$ with $\{ p_{n_1, n_1-2N'} (1,t)  \}$ and $\{ q_i \}$ with the normalized $\{ t^{-2N'^2} \}$, we have
\begin{eqnarray}
0&\leq&-\sum_{N'=0}^{\lfloor \frac{n_1}{2} \rfloor} \alpha t^{-2N'^2 -2(n-n_1)N' } \log\bigl[ \alpha t^{-2N'^2 -2(n-n_1)N' } \bigr]\nonumber\\
&<&-\sum_{N'=0}^{\lfloor \frac{n_1}{2} \rfloor} \alpha t^{-2N'^2 -2(n-n_1)N' } \log\Bigl[t^{-2N'^2}\Bigl( \sum_{M'=0}^{\lfloor \frac{n_1}{2} \rfloor}  t^{-2M'^2} \Bigr)^{-1}   \Bigr] . 
\label{quu}
\end{eqnarray}
By using Theorem 1, the normalization condition, and $n_1 \leq n$, the equation (\ref{quu}) is upper bounded as
\begin{eqnarray}
&\,&-\sum_{N'=0}^{\lfloor \frac{n_1}{2} \rfloor} \alpha t^{-2N'^2 -2(n-n_1)N' } \log\Bigl[t^{-2N'^2}\Bigl( \sum_{M'=0}^{\lfloor \frac{n_1}{2} \rfloor}  t^{-2M'^2} \Bigr)^{-1}   \Bigr] \nonumber\\
&\,&<\log\Big[ \sum_{M'=0}^{\infty}  t^{-2M'^2} \Big]+2C(t)^2 \log t\sum_{N'=0}^{\infty} N'^2 t^{-2N'^2}\, =: \, D_2(t)+D_3(t).
\end{eqnarray} 
By using the ratio test \cite{ratiotest}, it is easy to check that $D_1(s,t)$, $D_2(t)$ and $D_3(t)$ converge when $t>1$. The estimates of these $n$ independent constants are given in appendix B.

From the above, we see that the entanglement entropy is bounded as
\begin{eqnarray}
n_1 \log s +D_1(s,t)\leq S_{n_1}(s,t)<n_1 \log s +D_2(t)+D_3(t). 
\label{eq:EEbound}
\end{eqnarray}
Therefore, $S_{n_1}(s,t)=n_1 \log s + \Or(1)$, which means that the ground state obeys an area law when $s=1$, while it obeys a volume law when $s>1$. When we substitute $n$ for $n_1$, we have $S_{n}(s,t)=n \log s + \Or(1)$, which is consistent with the previous result in \cite{qfred}. Because $C(t) \rightarrow 1$ when $t\rightarrow \infty$, we can easily check that $D_1(s,t)\rightarrow 0$, $D_2(t)\rightarrow 0$, $D_3(t)\rightarrow 0$ and hence $S_{n_1}(s,t)\rightarrow n_1\log s$ in this limit. This corresponds to the fact that only the superposition of highest-area paths contributes in the limit $t\rightarrow \infty$. The volume-law scaling of $S_{n}(s>1,t)$ can be easily understood from this extreme limit. In this limit, the ground state takes the form 
\begin{equation}
|{\rm GS}\rangle = \frac{1}{\sqrt{s^n}} \sum_{\{ c_1, ..., c_n \}} 
|\uparrow^{c_1}_1 \uparrow^{c_2}_2 \cdots \uparrow^{c_n}_n 
\downarrow^{c_n}_{n+1} \cdots \downarrow^{c_2}_{2n-1} \downarrow^{c_1}_{2n} \rangle.
\end{equation}
The non-local correlation between up and down steps at the same height forces their colors to be the same in each state in the sum. Therefore, the states at $j$ and $2n-j+1$ in $|{\rm GS}\rangle$ form an entangled pair of $s$ levels, whose entanglement entropy is $\log s$. This is the source of the volume-law entanglement in the ground state of the model with $s>1$ and $t \gg 1$. We expect that this interpretation qualitatively explains the rigorous result Eq. (\ref{eq:EEbound}), which is valid for all $t>1$.

\subsection{R\'enyi entropy}
Here we calculate the R\'enyi entropy defined by
\begin{eqnarray}
S_{n_1}^{\kappa}(s,t)=\frac{1}{1-\kappa} \log \bigl[ {\rm Tr}(\rho^{\kappa}) \bigr]\,\,\,\,\,\,\,(\kappa\geq0,\,\kappa\neq 1).\label{renyi}
\end{eqnarray}
From the density matrix (\ref{dondon}), the R\'enyi entropy is
\begin{eqnarray}
S_{n_1}^{\kappa}(s,t)&=&\frac{1}{1-\kappa} \log \Bigl[ \sum_{N'=0}^{\lfloor \frac{n_1}{2} \rfloor} \alpha^{\kappa}s^{(2N'-n_1)\kappa}\,t^{-2\kappa N'^2 -2(n-n_1)\kappa N'}s^{n_1 -2N'} \Bigr]\nonumber\\
&=&n_1\log s +\frac{1}{1-\kappa}\log \Bigl[ \sum_{N'=0}^{\lfloor \frac{n_1}{2} \rfloor} \alpha^{\kappa}s^{-2(1-\kappa)N'} t^{-2\kappa N'^2 -2(n-n_1)\kappa N'}\Bigr]. \label{renyientropy111}
\end{eqnarray}
By using Theorem 1 and the fact that $n_1\leq n$, we have
\begin{eqnarray}
&&\sum_{N'=0}^{\lfloor \frac{n_1}{2} \rfloor}\alpha^{\kappa} s^{-2(1-\kappa)N'} t^{-2\kappa N'^2 -2(n-n_1)\kappa N'}\nonumber\\
&&<C(t)^{2\kappa}\sum_{N'=0}^{\infty}s^{-2(1-\kappa)N'} t^{-2\kappa N'^2}
=: C(t)^{2\kappa}D_4(s,t,\kappa).\label{renyientropy112}
\end{eqnarray}
By using the ratio test \cite{ratiotest}, one can easily check that $D_4(s,t,\kappa)$ converges when $t>1$ and $\kappa>0$. Note that $D_4(s,t,\kappa)\rightarrow1$ when $t\rightarrow\infty$. In the following, we discuss separately the cases (i) $\kappa=0$, (ii) $0<\kappa<1$, and (iii) $1<\kappa$.\\
(i) $\kappa=0$\\
\ \ \ \ \ \ By definition (\ref{renyi}), the R\'enyi entropy for $\kappa=0$ is $S_{n_1}^{0}(s,t)=\log \chi_{n_1}(s)$. Therefore, by using equations (\ref{schmidtrank1}) and (\ref{schmidtrank2}), we have
\begin{equation}
S_{n_1}^{0}(s,t)=\left \{
\begin{array}{l}
\log\Bigl(\left\lfloor \frac{n_1}{2} \right\rfloor +1\Bigr)\ \ \ \ \ \ \ \ \ \ \ \ \ \ \ \ \ \ \ \ \,{\rm for}\ \ s=1, \\
n_1\log s + \log \Bigl[ \frac{1-s^{-2(\lfloor \frac{n_1}{2} \rfloor+1)}}{1-s^{-2}} \Bigr]\ \ \ \ {\rm for}\ \ s>1.
\end{array}
\right.
\end{equation}
(ii) $0<\kappa<1$\\
\ \ \ \ \ \ We consider the fact that the R\'enyi entropy $S_{n_1}^{\kappa}(s,t)$ is non-increasing in $\kappa$ \cite{monotonicity}. By using equations (\ref{eq:EEbound}), (\ref{renyientropy111}) and (\ref{renyientropy112}), the R\'enyi entropy $S_{n_1}^{\kappa}(s,t)$ for $0<\kappa<1$ is bounded from above and below as follows:
\begin{eqnarray}
n_1\log s &+&D_1(s,t)\leq S_{n_1}(s,t)\leq S_{n_1}^{\kappa}(s,t)\leq S_{n_1}^{\kappa/2}(s,t)\nonumber\\
&<&n_1\log s +\frac{2\kappa}{2-\kappa}\log C(t)+\frac{2}{2-\kappa}\log D_4(s,t,\kappa/2).
\end{eqnarray}
Therefore, the R\'enyi entropy is $S_{n_1}^{\kappa}(s,t)=n_1\log s+\Or(1)$. In particular, in the limit $t\rightarrow \infty$, we have $S_{n_1}^{\kappa}(s,t) \rightarrow n_1\log s$.\\
(iii) $1<\kappa$\\
\ \ \ \ \ \ By considering the monotonicity of $S_{n_1}^{\kappa}(s,t)$ in $\kappa$ \cite{monotonicity} and using equations (\ref{eq:EEbound}), (\ref{renyientropy111}) and (\ref{renyientropy112}), we can bound the R\'enyi entropy from above and below as follows:
\begin{eqnarray}
n_1\log s &+&D_2(t)+D_3(t)>S_{n_1}(s,t)\geq S_{n_1}^{\kappa}(s,t)\geq S_{n_1}^{2\kappa}(s,t)\nonumber\\
&>&n_1\log s +\frac{4\kappa}{1-2\kappa}\log C(t)+\frac{1}{1-2\kappa}\log D_4(s,t,2\kappa).
\end{eqnarray}
Therefore, we have $S_{n_1}^{\kappa}(s,t)=n_1\log s+\Or(1)$ and $\displaystyle \lim_{t\to\infty}S_{n_1}^{\kappa}(s,t)=n_1\log s$.

\subsection{Entanglement spectrum}
The entanglement spectrum is the eigenvalue spectrum of the entanglement Hamiltonian $H_{\rm E}$ defined by
\begin{eqnarray}
\rho= e^{-\beta H_{\rm E}},
\end{eqnarray}
where $\beta$ is the inverse temperature~\cite{eham}. In the following, we set $\beta =1$ for simplicity. From Eq. (\ref{dondon}), we have
\begin{eqnarray}
\!\!\!\!\!\!\!\!\!\!
H_{\rm E} &=&-\log \rho \nonumber\\
&=&-\sum_{N'=0}^{\lfloor \frac{n_1}{2} \rfloor} \log \Bigl[ p_{n_1,n_1 -2N'}(s,t) \Bigr] \sum_{x\in\{\uparrow^1,\uparrow^2,\cdots,\uparrow^s\}^{n_1 -2N'}}\ket{\hat{C}_{0,n_1 -2N',x}}\bra{\hat{C}_{0,n_1 -2N',x}}, \nonumber \\
\end{eqnarray}
from which we can read off the eigenvalues of $H_{\rm E}$ as
\begin{equation}
E_{n_1, N'} = -\log \Bigl[ p_{n_1,n_1 -2N'}(s,t) \Bigr].
\end{equation}
Note that each $E_{n_1,N'}$ is $s^{n_1 -2N'}$-fold degenerate. By using Theorem 1, we can bound $E_{n_1,N'}$ from above and below as
\begin{eqnarray}
\!\!\!\!\!\!\!\!\!\!
E_{n_1,N'}<(2N'^2 +2N' n -2N' n_1)\log t + (n_1 -2N')\log s +\log C(t),
\end{eqnarray}
\begin{eqnarray}
\!\!\!\!\!\!\!\!\!\!
E_{n_1,N'}>(2N'^2 +2N' n -2N' n_1)\log t + (n_1 -2N')\log s -2\log C(t).
\end{eqnarray}
Now we introduce the factor $\gamma$ defined by $\gamma := n_1/n$ with $0<\gamma \leq 1$. 
Then we can rewrite the above result as
\begin{eqnarray}
E_{n_1,N'}=\bigl[2N'^2 +2N' n (1-\gamma)\bigr]\log t + (\gamma n -2N')\log s +\Or(1).
\end{eqnarray}
When we take the thermodynamic limit with keeping $\gamma$ constant, the entanglement spectrum {\it per site} in the low energy regime, namely $0\leq N'\ll n$, is
\begin{eqnarray}
\frac{E_{n_1,N'}}{2n}\approx N'(1-\gamma )\log t +\frac{\gamma}{2}\log s.
\end{eqnarray} 
Therefore, unless $\gamma=1$, the entanglement spectrum per site is approximately equidistant at low energies, which is reminiscent of the spectrum of the harmonic oscillator.

\section{Conclusion}
In this paper, we have studied the properties of the deformed Fredkin spin chain, particularly when the deformation parameter $t>1$. 
First, we proved that the finite-size gap is exponentially small in the system size. Second, we showed that the magnetization in the ground state is along the $z$-direction, and gave a rough estimate for $\braket{s^z}$. 
Third, we evaluated some entanglement-related quantities. In particular, we derive upper and lower bounds for the entanglement entropy and show that the ground state of the colorless ($s=1$) model obeys an area law, while that of the colored ($s>1$) model obeys a volume law.  We also studied the entanglement spectrum per site and found that its low-energy spectrum forms an equidistant spectrum. 
These results advanced our understanding of the deformed Fredkin spin chain.

However, many problems still remain to be solved. Examples include two-point functions such as $\braket{s^x s^z}$ and block entanglement. Furthermore, the properties of the model in the parameter region $t<1$ should be addressed in future studies.  

\ack
The authors thank Israel Klich and Vladimir E. Korepin for valuable discussions. After completion of this work, we learned that similar results have been obtained by Zhang and Klich~\cite{Zhang_prep}. HK was supported in part by JSPS KAKENHI Grant No. JP15K17719 and No. JP16H00985.

\appendix

\section*{Appendix. Estimates of $D_1(s,t)$, $D_2(t)$ and $D_3(t)$}
$D_2(t)$ can be bounded from above as
\begin{eqnarray}
D_2(t)&=&\log\Big[ \sum_{M'=0}^{\infty}  t^{-2M'^2} \Big]\nonumber\\
 &<&\log\Big[ 1+\int_{0}^{\infty} dx\, t^{-2x^2} \Big] =\log\Bigl[ 1+\frac{1}{2}\sqrt{\frac{\pi}{2\log t}} \, \Bigr].\nonumber
\end{eqnarray}
$D_1(s,t)$ and $D_3(t)$ are roughly estimated as
\begin{eqnarray}
D_1(s,t)&=&-2C(t)^2\log s \sum_{N'=0}^{\infty} N' t^{-2N'^2}\nonumber\\
&\sim&-2C(t)^2\log s \int_{0}^{\infty}dx\, x t^{-2x^2} =-\frac{C(t)^2 \log s}{2 \log t},\nonumber
\end{eqnarray}
\begin{eqnarray}
D_3(t)&=&2C(t)^2 \log t\sum_{N'=0}^{\infty} N'^2 t^{-2N'^2}\nonumber\\
&\sim& 2C(t)^2 \log t\int_{0}^{\infty} dx\, x^2 t^{-2x^2}=\frac{C(t)^2}{4}\sqrt{\frac{\pi}{2\log t}}.\nonumber
\end{eqnarray}

\end{document}